# Zero Focal Shift in High Numerical Aperture Focusing of a Gaussian Laser Beam through Multiple Dielectric Interfaces

Ali Mahmoudi

a.mahmoudi@qom.ac.ir  &  amahmodi@yahoo.com

Laboratory of Optical Microscopy, Department of Physics, Faculty of sciences, University of Qom, Qom, Iran

Abstract- In this paper, focal shift of a Guassian laser beam tightly focused through several planar dielectric interfaces is discussed. It is shown that the focal shift and focus point can be changed by variation of refractive index mismatch and this shift could be positive or negative, and it is proportional to refractive index mismatch between coverglass and immersion medium. It is shown also that for every immersion medium(coverglass), focal shift could be removed for a continuous range of depths by choosing a coverglass (immersion medium) with appropriate refractive index.

**1-Introduction**- When a laser beam is focused through a dielectric interface between two homogenous dielectric media, induced spherical aberration at interface results in focal shift which means that the actual axial position of the focal peak is shifted from focus given by geometrical optics [1]. In optical microscopy, location or depth of geometrical focus is determined by the actual distance travelled by lens (microscope objective); this is sometimes called probe depth. Thus one could not know the real distance or depth of laser focal spot, and this results in problems in applications such as laser tweezers, optical microscopy, confocal fluorscence microscopy and data storage systems. For example, in confocal fluorescence microscopy, location of a single fluorophore could not be determined exactly [2]. Normally, in confocal fluorescence microscopy or optical trapping, laser light is focused inside a glass chamber containing aqueous sample medium. So, there are several planar interfaces in laser light path. The media are objective (glass, $n_{obj}$), immersion medium (oil, air, etc., $n_{im}$), coverglass (normally, with same refractive index as objective material, $n_g$), and sample medium(water or another material, $n_s$). If these media have different refractive indices, spherical aberrations have different sources. These aberrations result in a complex focus shift, degradation of focal spot, and axial broadening of point spread function that reduces axial resolution. Different techniques proposed for compensation of these aberrations [3-5]. Recently it is shown experimentally that these aberrations could be compensated or corrected using phase only spatial light modulators [6]. Focusing light through planar interface was studied by Peter Török et.al. Using angular spectrum of plane waves and Debye approximations for high aperture focusing, Török et al. extended the Richards and Wolf model [7], and derive an expression for electrical field of focused light through a planar interface between two mismatched media [8-10]. Using an extension of their analysis, where there are more than one mismatched interfaces in laser pathway, we have shown that the optimal depth in optical tweezers; where optical trap has maximum stiffness and spherical aberration is minimized, can be changed using immersion oils with different refractive indices [11]. This effect was shown experimentally by S.N.S.Reihani et.al.[12]. The exact focus location is also of very importance in application in silicon (Si) integrated circuit analysis and in optical data storage, too [13]. In these applications, high resolution and sensitivity imaging of subsurface features is needed [14].

This paper is organized as follows; section 2 describes intensity distribution around the focal point in presence of several planar interfaces between homogenous media. In section 3, role of refractive index mismatch between these media in focal shift is calculated and discussed and it is shown that using a variable refractive index material as coverglass or immersion medium, one can remove focal shifts in a continuous range of probe depths, and section 4 presents the conclusions.

**2- Intensity distribution inside the sample medium around focus inside sample medium**

When a linearly polarized (monochromatic) laser beam is focused through an aplanatic objective lens into a homogeneous medium, It can be shown that the electric field at a given point P around focus (the origin located at the Gaussian focus center; defined by point "O" in Fig. 1) can be written as[7] :

$$\vec{E} = -\frac{ik}{2\pi} \iint_\Omega \frac{\vec{a}(s_x, s_y)}{s_z} \exp[ik[\Phi(s_x, s_y) + \hat{s}.\vec{r_P}]] ds_x ds_y \qquad (1)$$

Where k is wavenumber in medium, $\hat{s} = s_x \hat{i} + s_y \hat{j} + s_z \hat{k}$ is a unit vector along a typical ray or k-vector of one of the constituent plane waves, $\Omega$ is the solid angle formed by all rays emerging the lens (defined by N.A. of the objective lens), $\vec{r_P}$ is the position vector of point $P$, $\Phi$ is the spherical aberration function introduced by objective itself, and $\vec{a}$ is electric strength vector at the entrance aperture of objective. The equation for magnetic field is similar. In presence of a planar interface located at $z = -Z_I$ between two different isotropic media (with refractive indices $n_1$ and $n_2$ ) in image space, and assuming that no aberration is induced by objective lens ($\Phi = 0$), it is shown that the electric field at the interface (inside second medium) can be written as [8]:

$$\vec{E_2}(x, y, -Z_I) = -\frac{ik_1}{2\pi} \iint_\Omega T^{1\to 2} W_e(\hat{s}_1) \exp[ik_1(s_{1x}x + s_{1y}y - s_{1z}Z_I)] ds_1 ds_2 \qquad (2)$$

Where $k_1$ is wavenumber in medium 1, $W_e = \frac{\vec{a}(s_{1x}, s_{1y})}{s_{1z}}$ , and $T^{1\to 2}$ is a matrix (operator) that describes the change of electric filed of a typical ray on passing through the interface. This matrix is a function of angle of incidence and refractive indices of two media and contains Fresnel transmission coefficients. Repeating this analysis, one can extend equation 2 to find an expression for electric field inside sample medium in presence of several planar interfaces. Figure 1 shows a schematic view of focusing laser light using a high numerical aperture lens inside a sample medium. Solid lines show real optical path for two typical rays, where dashed lines are for homogenous case. Using this extension it can be shown that distribution of the electric field around focus in spherical coordinates can be written as [11]:

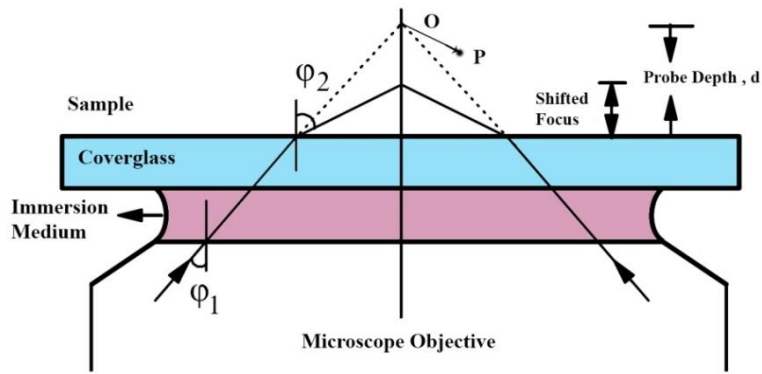

**Figure 1: A typical ray originating from objective travels inside sample medium. Guassian (geometrical) focus located at O. The actual focus is shifted toward coverglass. Here, refractive index of coverglass is higher than refractive index of the sample medium, $n_g > n_s$ and $n_{obj} = n_{im} = n_g$.**

$$\vec{E_4}(x,y,z)$$
$$= \int_0^\alpha \int_0^{2\pi} C\, \vec{e}_s \exp[ik_1 n1(\sum_{m=2}^4 d_m \cos\varphi_1 - \sum_{m=2}^4 n_m d_m \cos\varphi_m)] \exp[in_4 k_1 z \cos\varphi_4] \exp[in_1 k_1 \sin\varphi_1 (x\cos\theta$$
$$+ y\sin\theta)] \sin\varphi_1 \cos\varphi_1^{\frac{1}{2}} d\theta d\phi_1 \qquad (3)$$

Where C is an amplitude constant, and $\vec{e}_s$ is electrical strength vector in sample medium [11]. $d_m$ is the distance travelled by laser light in m-th medium, so that $d_2$ is thickness of the immersion layer, $d_3$ is thickness of coverglass, and $d_4 = d$ is probe depth, or depth inside sample medium. $\varphi_m$ and $n_m$ are angle of incidence of a typical ray in m-th medium, and refractive index of m_th medium, respectively. $\alpha$ is the maximum angle determined by numerical aperture (N.A.) of the objective lens. If the system has axial symmetry around optical axis, integration over $\theta$ can be performed analytically and electric field vector will be written as:

$$E_{4x}(x,y,z) = B \int_0^\alpha A(\varphi) \exp[ik_1 n1(\sum_{m=2}^4 d_m \cos\varphi_1$$
$$- \sum_{m=2}^4 n_m d_m \cos\varphi_m)] \exp[in_4 k_1 z \cos\varphi_4] \sin\varphi_1 \cos\varphi_1^{\frac{1}{2}} [-\prod_1^N \tau_{pj} \cos\varphi_4$$
$$+ \prod_1^N \tau_{sj}][J_0(n_1 k_1 r \sin\varphi_1) - J_2(n_1 k_1 r \sin\varphi_1) \frac{x^2 - y^2}{r^2}] d\phi_1$$

$$E_{4y}(x,y,z) = B \int_0^\alpha A(\varphi) \exp[ik_1 n1(\sum_{m=2}^4 d_m \cos\varphi_1$$
$$- \sum_{m=2}^4 n_m d_m \cos\varphi_m)] \exp[in_4 k_1 z \cos\varphi_4] \sin\varphi_1 \cos\varphi_1^{\frac{1}{2}} [\prod_1^N \tau_{pj} \cos\varphi_4$$
$$- \prod_1^N \tau_{sj}][J_2(n_1 k_1 r \sin\varphi_1) \frac{2xy}{r^2}] d\phi_1 \qquad (4)$$

$$E_{4z}(x,y,z) = -2iB \int_0^\alpha A(\varphi) \exp[ik_1 n1(\sum_{m=2}^4 d_m \cos\varphi_1$$
$$- \sum_{m=2}^4 n_m d_m \cos\varphi_m)] \exp[in_4 k_1 z \cos\varphi_4] \sin\varphi_1 \cos\varphi_1^{\frac{1}{2}} [\prod_1^N \tau_{pj} \sin\varphi_4$$
$$+ \prod_1^N \tau_{sj}][J_1(n_1 k_1 r \sin\varphi_1) \frac{x}{r}] d\phi_1$$

Where B is constant, A is a function that describes amplitude distribution (laser transverse mode) over entrance aperture of the objective lens. $\tau_{pj}$ and $\tau_{sj}$ are Fresnel transmission coefficients for p and s polarizations at the j-th interface. r is distance to optical axis and is equal to $\sqrt{x^2 + y^2}$ and $J_0$ and $J_1$ are Bessel functions of first kind of first and second orders, respectively.

**3- Role of refractive index of objective immersion medium and coverglass in focal shift**

Numerical computation of equation (4) gives the intensity distribution around focal point. Setting (x,y,z)=(0,0,z), intensity distribution on optical axis is computed from Eq. 4. Consider a linearly polarized TEM00 Gaussian laser beam ($\lambda = 1064nm$) that is focused using a high numerical aperture microscope objective (N.A=1.3 and focal length f) , into a sample medium through a set of planar interfaces between different media. Then $A(\varphi) = \exp[-f^2 sin^2\varphi_1/w_0^2]$, where $w_0$ is beam waist of laser beam on lens aperture.

Figure 2 shows axial intensity distribution around focal point, probe depth is fixed and equal to 10 μm, when the image space is homogenous (no mismatched interface), focal shift is zero and axial position of focal peak is equal to probe depth. When there is one mismatched interface in the image space (coverglass-sample medium), negative focal shift occurs, in this case immersion medium with a refractive index below 1.52 (=$n_g$) results in larger focal shift. On the other hand, positive focal shift is clearly seen when $n_{im}$>1.52. For computing focal shifts, one needs only to calculate axial position of on-axis intensity peak ($Z_{Max}$).

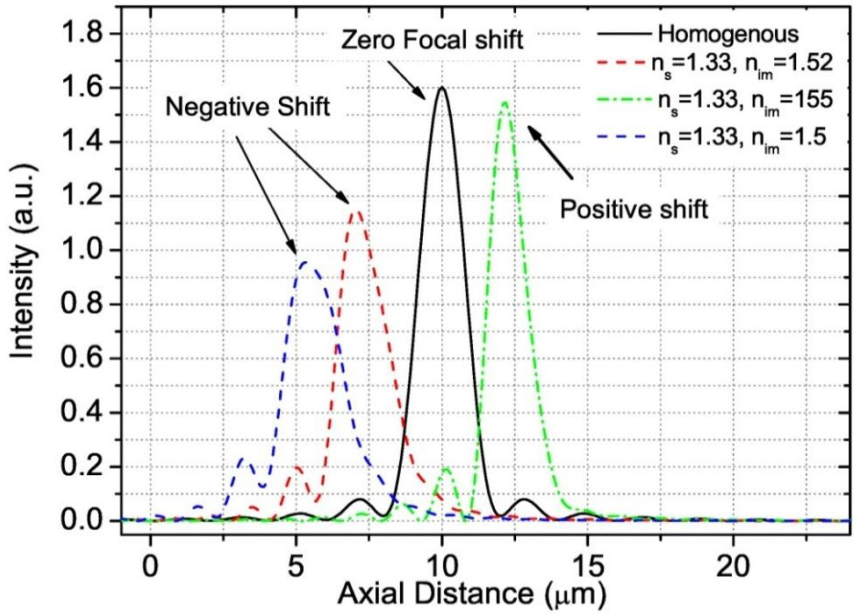

Figure 2: positive and negative focal shifts occur in presence of refractive index mismatch between different media in the image space. Probe depth is fixed and equal to $10\mu m$ . When the image space is homogenous, there is no focal shift.

If the probe depth is d, then focal shift will be  ($|\Delta f| = Z_{Max} - d$). In the subsequent sections, the effect of refractive index of coverglass and immersion medium on focal shift is discussed.

Figure 3 (a) shows absolute value of focal shift ($|\Delta f|$) for constant probe depth (d) as a function of refractive index of immersion medium for different coverslip glasses (different $n_g$) with same thicknesses (=$170\mu m$). The probe depth is fixed and equal to 10 $\mu m$. The sample medium is assumed to be water ($n_s = 1.33$). As it can be seen clearly, the focal shift changes linearly with refractive index of immersion medium, first it decreases ($Z_{Max} - d$ is negative), then reaches a minimum (zero) and then it increases linearly ($Z_{Max} - d$ is positive). For every coverglass, there is an immersion medium (with refractive index $n_0$) which minimizes the focal shift ($Z_{Max} - d = 0$). Thus, changing refractive index of the immersion medium gradually, results in gradual variation of the focal shift. Liquid crystal materials with variable refractive index were proposed for variable focusing lenses and are available now [15,16]. So, using a transparent variable refractive index material as immersion medium or coverglass, it is possible to have a microscope objective with variable focus shift or a variable focal point (length). If this change in refractive index is fast, then it is possible to switch between focal points. This would be simple and inexpensive. It is worth mentioning that by measuring a focal shift for an unknown material, one can measure its refractive index. It should be mentioned that, as it is seen in figure 3, around $n_0$, every 0.1 change in refractive index of the immersion medium, results in about $2\mu m$  increase in the focal shift. The shifts are positive for $n_{im} > n_0$  and negative when  $n_{im} < n_0$. Figure 3 (b) shows focal shift as a function of refractive index of coverglass  for different immersion media. Probe depth is fixed and $10\mu m$, thickness of coverglass is 170 $\mu m$. Again, it can be seen clearly that for every immersion medium, as $n_g$ increases, focal shift is decreased ($Z_{Max} - d$ is negative) linearly, then it takes a minimum (zero focal shift for $n_g = n_0$), and again it increases linearly. It should be mentioned that around $n_0$, every 0.1 change in refractive index of the coverglass, results in about $1.6\mu m$ increase in the focal shift. These shifts are positive if $n_g > n_0$ and negative if $n_{im} < n_0$.

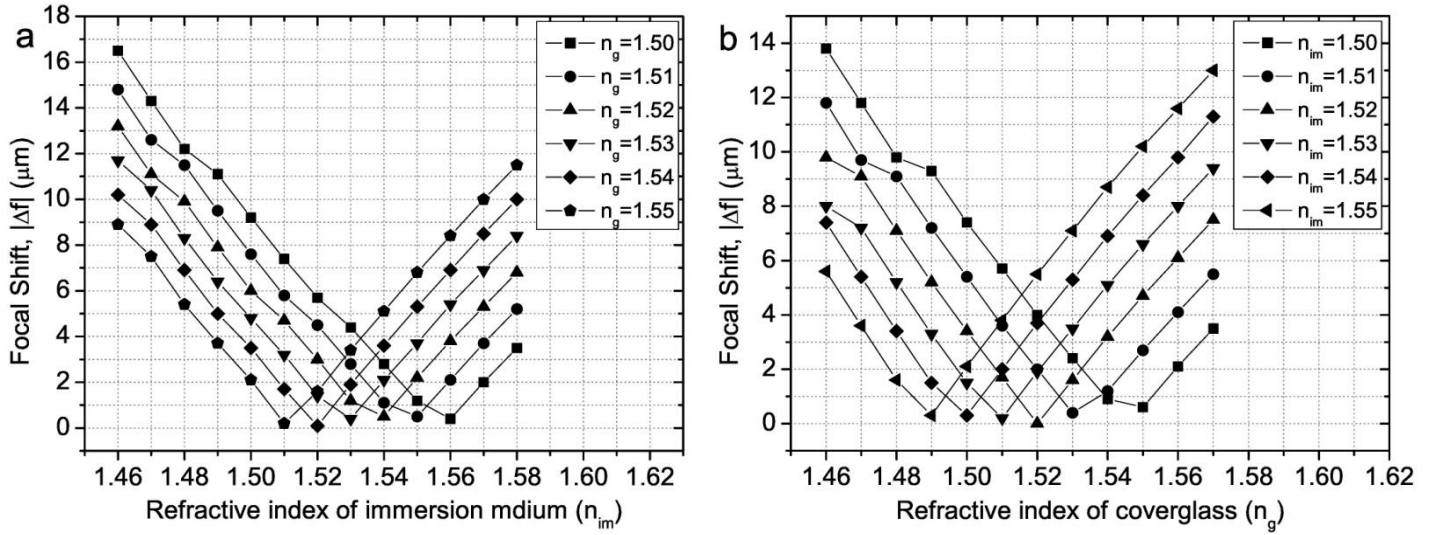

Figure 3: a: Focal shift versus refractive index of immersion medium for different values of $n_g$. Probe depth is kept fixed and equal to 10 micrometers. b: Focal shift versus refractive index of coverglass for different values of $n_{im}$. Probe depth is kept fixed and equal to 10 micrometers.

In figure 4 (a) probe depth with zero focal shift (D₀) is shown versus refractive index of immersion medium ($n_{im}$). Coverglass with fixed refractive index is considered here ($n_g = 1.52$). It is clearly seen that D₀ changes linearly with n_{im}, and focal shift could be removed for a continous range of depths using a transparent material with variable refractive index as immersion medium. In figure 4 (b) same calculations is shown for D₀ as a function of $n_g$, where an immersion medium with $n_{im} = 1.52$ is used. Again it is seen that a coverglass with variable refractive index could result in zero focal shift for a continous range of probe depths. It is very important that in both cases, D₀ changes linearly with refractive index of immersion medium or coverglass and hence with refractive index mismatch.

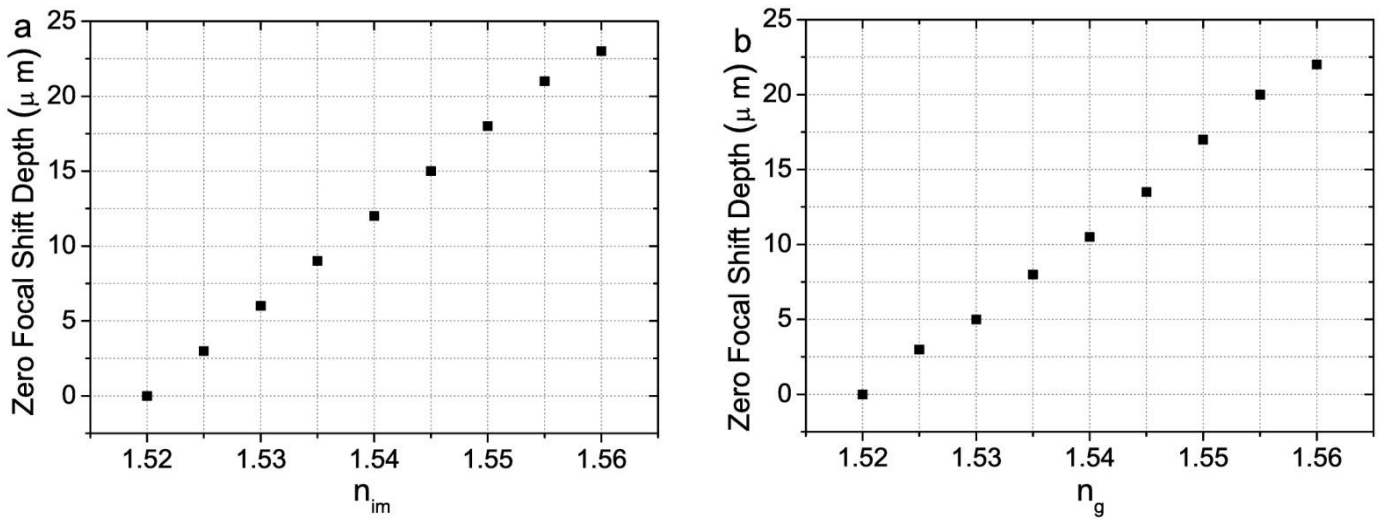

Figure 4: (a) Zero focal shift depth, D₀ versus n_{im}, there is a depth with minimum focal shift for every immersion medium. (b): Zero focal shift depth D₀ as a function of n_g. In both cases, D₀ changes linearly with refractive index.

**Conclusions**

The role of refractive index mismatch between several media with planar interfaces in focal shift has been discussed. It is shown that, using an immersion medium (coverglass) with variable refractive index, focal shift can be changed or removed. For an immersion medium (coverglass), it is shown numerically that every 0.1 change in refractive index mismatch (immersion medium-coverglass), results in about $2\mu m$ ($1.6\mu m$) increase in the focal shift. It is shown also that zero focal shifts can be achieved in a continuous range of depths. So, one can design an inexpensive microscope without focal shift using a material with variable refractive index as immersion medium or coverglass.